\documentclass[prd,aps,graphics]{revtex4}
\usepackage{epsfig}

\newcommand{\be}{\begin{equation}}
\newcommand{\ee}{\end{equation}}
\newcommand{\bea}{\begin{eqnarray}}
\newcommand{\eea}{\end{eqnarray}}
\newcommand{\ket}{ | 0 \rangle }
\newcommand{\bra}{ \langle 0 | }
\newcommand{\kete}{ | 0 \rangle_{\eta_2} }
\newcommand{\brae}{ {}_{\eta_1} \! \langle 0 | }

\def\vx{\vec{x}}
\def\vk{\vec{k}}
\def\vp{\vec{p}}

\def\e1{\eta_1}
\def\e2{\eta_2}
\def\u1{\u^{(1)}}
\def\u1*{\u^{(1)*}}
\def\k1{\k_1}
\def\k2{\k_2}
\def\k3{\k_3}

\begin{document}
\title{Radiative corrections to scalar masses in de Sitter space}

\author{Tristan Brunier and Francis Bernardeau}
 \email{brunier@spht.saclay.cea.fr}
 \email{fbernard@spht.saclay.cea.fr}
 \affiliation{Service de Physique Th{\'e}orique,
         CEA/DSM/SPhT, Unit{\'e} de recherche associ{\'e}e au CNRS, CEA/Saclay
         91191 Gif-sur-Yvette c{\'e}dex}

\author{Jean-Philippe Uzan}
 \email{uzan@th.u-psud.fr}
 \affiliation{Laboratoire de Physique Th{\'e}orique, CNRS--UMR 8627,
              B{\^a}t. 210, Universit{\'e} Paris XI,
              F-91405 Orsay Cedex, France,\\
              Institut d'Astrophysique de Paris, GReCO,
              CNRS-FRE 2435, 98 bis, Bd Arago, 75014 Paris, France.}

\vskip 0.15cm

\begin{abstract}
We compute the radiative corrections to the mass of a test boson field in an inflating space-time. The calculations are carried out in case of a boson part of a supersymmetric chiral multiplet. We show that its mass is preserved up to logarithmic divergences both in ultraviolet and infrared domains. Consequences of these results for inflationary models are discussed.
\end{abstract}

 \vskip2pc

\maketitle
%%%%%%%%%%%%%%%%%%%%%%%%%%%%%%%%%%%%%%%%%%%%%%%%%%%%%%%%%%%%%%%%%%%
\section{Motivation}

Quantum field theory in an inflating universe is thought to be the playground of the physical processes that took place during the early Universe \cite{inflation}. However little has yet been explicitly computed for self interacting fields \cite{birrel}. Any light bosonic fields with non minimal coupling that exist during an inflationary phase is bound to produce, together with the inflaton field, superhorizon fluctuations that eventually might be visible in mechanisms such as isocurvature modes generation from multi-field inflation
\cite{isocourbure,isobs1,isobs2,NGiso},
curvaton models \cite{curvaton,BMR}, bent trajectory inflationary models
\cite{BU1,BU2} or modulated fluctuations \cite{BU3}. If such fields are
self interacting, for instance with a quartic potential, such
fluctuations might develop significant non-Gaussian features that
in turn could be detected (see ref.\cite{matarrese} for a review).
This would be only possible however if
the radiative corrections do not render the particle too massive,
e.g. with a mass larger than the Hubble constant $H$,
to develop any fluctuations at all. One question then raised
by phenomenological investigations of this physics is whether the mass
of a self interacting bosonic field can be protected against
radiative corrections.

Of course one expects such a theory to be renormalizable but it implies that
the fundamental bare theory has to be fine-tuned.
The radiative correction to the mass of a self interacting boson
should indeed be naively of the order of
$\delta m^2 \sim \lambda\,M_{\rm Pl}^2$
($\lambda$ being the coupling constant). This question has been investigated
in \cite{Linde} in the case of the inflaton field. In this case $\lambda$ is generically small since, for the inflaton field $\lambda\,M_{\rm Pl}^2$ has to be of the order of $H^2$. In the case we are interested in,
however, we have no such constraints on $\lambda$ which can then be close to unity.
For such values of the coupling constant
the scalar field becomes too heavy
to develop significant fluctuations unless the cancellation is
precise over ten orders of magnitude
- since $H^2 / M^2_{\rm Pl} \sim 10^{-10}$ during inflation.

On the other hand light scalar fields are expected to be
associated with fermions as parts of super-multiplets in
supersymmetric theories - such as in $D$ or $F$-term inflation
models \cite{DFterm, Riotto}. From the Minkowskian behavior, one expects
the largest divergences coming from the fermionic and bosonic
loops to cancel out. The aim of this paper is precisely to compute
the radiative corrections to a test boson mass embedded in a
chiral multiplet and when it lives in an inflating universe. The
test model will be the Wess-Zumino Lagrangian in an expanding
universe, not necessarily assumed to be de Sitter. The background
evolution of the Universe is assumed to be driven by an other
sector of the theory.

We set up the formalism in section II. In section III we compute 
the two-point function to get the one loop effective masses of 
the fields in any spacially flat 
Friedman-Lema\^{i}tre-Robertson-Walker (FLRW) spacetime. The
ultraviolet and infrared behaviors are discussed in section IV. We
show in particular that the infrared divergences can be
apprehended in a classical approach. In the ultraviolet domain,
the correction to the mass is found to be logarithmically
divergent and proportional to the spacetime curvature and the
coupling constant. It shows that masses of light scalar fields do
not get a large contribution compared to the Hubble scale. Our
results are put into perspective in the conclusion.

\section{General setup}

\subsection{A toy model}

We aim at studying radiative mass corrections in a generic Friedmann-Lemaitre-Robertson-Walker background with flat spatial sections where the background is assumed to be driven by another sector of the theory. We will focuss on the following Lagrangian
\be
\label{LWZ}
{ \mathcal{L} } = \frac{1}{2} g^{\mu\nu} \varphi^* \nabla_{ \mu \nu } \varphi + \frac{1}{2} \overline{\psi} \left[ i \gamma^{ \alpha } V_{ \alpha } \, ^ {\mu} (x) \nabla_{\mu} - m \right] \psi - \frac{ \lambda }{ \sqrt{2} } \overline{\psi} ( \phi_1 - i \gamma^5 \phi_2 ) \psi - m \left | \varphi \right |^2 - m \lambda \left | \varphi \right |^2 ( \varphi + \varphi^* ) - \lambda^2 \left | \varphi \right |^4
\ee
where $ \varphi = 1 / \sqrt{2} ( \phi_1 + i \phi_2 ) $ is a complex scalar field, $\psi$ is a four-dimensional Majorana spinor and the $\gamma^\alpha$ are the flat Dirac matrices. We also defined
\be
\overline{\psi} = \psi^{\dagger} \gamma^0 \; , \; \gamma^5 = i \gamma^0 \gamma^1 \gamma^2 \gamma^3.
\ee
The kinetic term for fermions involves the Vierbein which are defined as
\be
V^{\alpha}\,_{\mu}(x)=\left(\frac{\partial \zeta^{\alpha}_{X}(x)}{\partial x^{\mu}}\right)_{X=x}
\ee
where the coordinates $\zeta^{\alpha}(x)$ correspond to a locally inertial frame while $x^{\mu}$ are the non-inertial coordinates. One can check that the Vierbein behave like vectors under Lorenz transformation (index $\alpha$) and under coordinate transformation (index $\mu$).
The metric $g_{\mu\nu}$ is related to the Minkowkian one $\eta_{\alpha\beta}$ through
\be
g_{\mu\nu}(x)=V^{\alpha}\,_{\mu}(x)V^{\beta}\,_{\nu}(x)\eta_{\alpha\beta}.
\ee
To obtain an action which is invariant under Lorentz group and coordinates change, the derivatives of a spinor have to be modified in the following way
\be
\nabla_{\alpha} = V_{ \alpha } \, ^{ \mu }(x) \nabla_{ \mu } = V_{\alpha}\,^{\mu}(x) ( \partial_{ \mu } + \Gamma_{ \mu } )
\ee
where the spin connexion is defined by
\be
\Gamma_{\mu}(x) = \frac{1}{2} \sigma^{ \alpha \beta } V_{\alpha} \, ^{\nu} (x) \partial_{\mu} V_{ \beta \nu } (x) ,
\ee
$\sigma^{\alpha\beta}$ being the spinorial representation of the Lorentz group generators given in terms of the Dirac matrices
\be
\sigma^{\alpha\beta}=\frac{i}{4}[\gamma^{\alpha},\gamma^{\beta}].
\ee

Since we will focus on spacially flat FLRW spacetimes,
the metric should be written in conformal time such as $g_{\mu\nu}=a^2(\eta)\eta_{\mu\nu}$.
The Vierbein are simply expressed as $V^{\alpha}\,_{\mu}(x)=a\delta^{\alpha}\,_{\mu}$ and
defining the conformal fields as
\bea
\widetilde{ \varphi } & = & a \varphi \\
\widetilde{ \psi } & = & a^{3/2} \psi
\eea
the Lagrangian takes the form
\be
{\mathcal{L}} = \frac{1}{a^4} \Big[ \frac{1}{2} \eta^{\mu\nu} \widetilde{\varphi}^* \partial_{\mu\nu} \widetilde{\varphi} +
\frac{1}{2} \overline{\widetilde{\psi}} \left( i \gamma^{\alpha} \partial_{\alpha} - m a \right) \widetilde{\psi} +
\frac{a''}{a} \widetilde{\varphi}^* \widetilde{\varphi} - \frac{\lambda}{\sqrt{2}} \overline{ \widetilde{\psi}}
( \widetilde{\phi}_1 - i \gamma^5 \widetilde{\phi}_2 ) \widetilde{\psi} -
 m^2 a^2 \left| \widetilde{\varphi} \right| ^2 - \lambda^2 \left| \widetilde{ \varphi} \right|^4 - m \lambda a \left| \widetilde{\varphi} \right|^2 ( \widetilde{ \varphi } + \widetilde{ \varphi }^* ) \Big].
\ee
Recalling that the scalar curvature is given by $ R = 6 a'' / a^3 $, the bosons formally acquire an effective time-dependent negative mass square proportional to the curvature
\be
\label{meff}
m^2_{\rm eff} = - \frac{1}{6} R < 0
\ee
while the fermions stay massless. As we shall see in the following, quantum corrections to the mass in the context of a cut-off regularisation lead to logarithmic divergences that are proportional to the curvature, as we should expect from (\ref{meff}).

One can note that this Lagrangian is deduced from a Wess-Zumino model with the superpotential
\be
W(\Phi) = \frac{1}{2} m^2 \Phi + \frac{1}{3} \lambda \Phi^3.
\ee

However, this is only a Wess-Zumino inspired toy model. Supersymmetric models in curved spacetime should be carefully dealt with since the parameter of a supersymmetric transformation should be space and time dependent \cite{SUGRA1,SUGRA2}. The background structure may forbid any coherent global supersymmetric approach.

\subsection{Quantizations of the fields}

As we are interested in quantum corrections to masses, we shall quantize the fields in a curved background.

The bosonic fields in any potential $U(\phi)$ obey the
Klein-Gordon equation which reads \be \label{KG1}
\frac{1}{\sqrt{-g}} \partial_{\mu} (\sqrt{-g} g^{\mu\nu}
\partial_{\nu} \phi) = \frac{ {\rm d} U(\phi) } { {\rm d} \phi }.
\ee Setting $\chi=a\phi$ and $V(\chi)=U(\chi/a)$, one gets the
equation of motion written in a spacially flat FLRW metric \be
\label{KG} \chi''-\Delta \chi = - a^4 \frac{ {\rm d} V }{ {\rm d}
\chi } + \frac{1}{6} R a^2 \chi \ee where $R$ is the scalar
curvature, $a$ is the scale factor and a prime stands for a
derivative with respect to the conformal time. The friction term
in (\ref{KG1}) coming from the kinetic term has been absorbed in
an effective mass term in (\ref{KG}). Since flat spatial sections
are considered, the free real scalar fields may be expanded on a
plane wave basis 
\bea
\phi(\vx,\eta) &=& \frac{1}{a(\eta)} \int \frac{ {\rm d}^3 \vk }{(2\pi)^{3/2}} \left[ u_k(\eta) {\rm e}^{i \vk \cdot \vx} a_{\vk} + u_k^*(\eta) {\rm e}^{-i \vk \cdot \vx} a_{\vk}^{\dagger} \right]\\
&=& \frac{1}{a(\eta)} \int \frac{ {\rm d}^3 \vk }{(2\pi)^{3/2}} \phi_{\vk}(\eta) {\rm e}^{i \vk \cdot \vx}.
\eea
where the operators $a_{\vk}$ and $a^{\dagger}_{\vk}$ are respectively the annihilation and creation operators and satisfy the equal time commutations
\be
\label{commutation}
\left[a_{\vk},a^{\dagger}_{\vk'}\right]=\delta(\vk-\vk') , \left[a_{\vk},a_{\vk'}\right]=0 , \left[a^{\dagger}_{\vk},a^{\dagger}_{\vk'}\right]=0.
\ee
The canonical quantization imposes a condition on the Wronskian normalisation
\be
\label{wronskien}
u_k(\eta) {u_k^*}'(\eta) - u_k^*(\eta) u_k'(\eta) = i.
\ee
In the following, we will define the Wightman function to be
\be
\label{Gk}
G_k(\eta,\eta') = u_k(\eta) u_k^*(\eta') ,
\ee
its expression in a given background depending on the choice of the vacuum.

In a FLRW background with flat spatial sections, the Dirac equation reads
\be
i \gamma^{\mu} \partial_{\mu} \left( a^{3/2} \psi \right) - m a \left( a^{3/2} \psi \right) = - \frac{ {\rm d} U }{ {\rm d} \overline{\psi} }.
\ee

In the massless case, a free spinor in a spacially flat FLRW spacetime is just conformal to a Minkowski spinor. In the following, for the sake of simplicity we shall only consider massless fermions. Moreover, we will consider Majorana spinors satisfying the conjugation relation
\be
\label{conjugaison}
\psi=C(\overline{\psi})^{T}
\ee
where C is the charge conjugation matrix that satisfies the property $C^{-1}=-C=C^T$.
Consequently the decomposition of a non-massive Majorana spinor on a plane wave basis is trivialy related to the Minkowskian decomposition and reads
\be
\psi(\vx,\eta) = \frac{1}{ a^{3/2} } \int \frac{ {\rm d}^{3} \vk }{ (2\pi)^{3/2} } \frac{1}{ \sqrt{2k} } \sum_{j=1}^{2} \left[ u^{(j)}(k) {\rm e}^{-i k \eta} {\rm e}^{+i \vec{k} \cdot \vec{x} } b^{j}_{\vec{k}} + v^{(j)}(k) {\rm e}^{+i k \eta} {\rm e}^{-i \vec{k} \cdot \vec{x}} { b^{j} }^{\dagger}_{\vec{k}} \right]
\ee
where  $k$ stands for $\| \vk\|$ and $u^{(j)}(k)$ et $v^{(j)}(k)$ are four-components spinors related by $u^{(j)}=C{(\bar{v}^{(j)})}^T$.
The fermionic operators commute with the bosonic ones and satisfy the anticommutation relations
\be
\label{anticommute}
\left\{b^j_{\vk},{b^{j'}}_{\vk'}^{\dagger}\right\}=\delta_{jj'}\delta(\vk-\vk') \;,\; \left\{b^j_{\vk},{b^{j'}}_{\vk'}\right\}=0 \;,\; \left\{{b^j}_{\vk}^{\dagger},{b^{j'}}_{\vk'}^{\dagger}\right\}=0.
\ee
The free vacuum is defined by the conditions
\be
\label{vacuumfermi}
a_{\vk} \ket = 0 \;\;b^{j}_{\vk} \ket = 0 \;\;\forall \;j,\vk.
\ee
As in Minkowski, the mode functions satisfy the completness relation
\be
\label{completness1}
\sum_{j=1}^{2} u^{(j)}_{\alpha}(k)\bar{u}^{(j)}_{\beta}(k)=k^{\mu}(\gamma_{\mu})_{\alpha\beta}=\sum_{j=1}^{2}\bar{v}^{(j)}_{\alpha}(k)v^{(j)}_{\beta}(k).
\ee

Using the relations (\ref{conjugaison}) and (\ref{completness1}), one also finds
\bea
\sum_{\alpha,\beta=0}^{4} \bra \overline{\psi}_{\alpha}(x') \overline{\psi}_{\beta}(x'') \ket \bra \psi_{\alpha}(x') \psi_{\beta}(x'') \ket & = & - \sum_{ \alpha , \beta =0 }^{4} \bra \overline{\psi} _{ \alpha } (x') \psi_{ \beta } (x'') \ket \bra \psi_{ \alpha } (x') \overline{\psi}_{ \beta } (x'') \ket \nonumber\\
&=& - \frac{1}{a^3(\eta')a^3(\eta'')}\int \frac{ {\rm d}^{3} \vp_1 {\rm d}^3 \vp_2}{ (2\pi)^6 } \Delta_{p_1,p_2}(\eta',\eta''){\rm e}^{i(\vp_1+\vp_2)\cdot(\vx'-\vx'')}
\label{Correlateur}
\eea
with
\be
\label{fermion3}
\Delta_{p_1,p_2}(\eta',\eta'')=\frac{\eta^{\mu\nu}{p_1}_{\mu}{p_2}_{\nu}}{4p_1p_2}{\rm e}^{-i(p_1+p_2)(\eta'-\eta'')}
\ee
where we set $p_i=\|\vp_i\|$.

We are now ready to calculate radiative mass corrections. In the next section, we set up the formalism we shall use to study one loop mass correction in Yukawa potentials and compare our results to the effect of a small mass.

\section{Mass correction}

\subsection{Quantum level formalism}

In order to study radiative mass corrections in a generic spacially flat FLRW background, we are interested in the calculation of the two-point function
\be
\brae \phi_{\vk_1}(\eta_1)U(\eta_1,\eta_2)\phi_{\vk_2}(\eta_2)\kete
\ee
where $\phi$ is a scalar field, $|0\rangle_{\eta_i}$ is the vacuum state at the time $\eta_i$ and $U(\eta_1,\eta_2)$ is an evolution operator that makes a state evolve from the conformal time $\eta_2$ to the conformal time $\eta_1$.

In a free field theory, this expression may be easily computed. For example in an Heisenberg picture, the vacuum state does not evolve and the form of the field depends only on the background and on the choice of the vacuum state. However this expression must be modified in an interacting field theory and the corrections can only be evaluated order by order in the coupling constants in a perturbation theory approach.

In an interaction picture, the states evolve under the action of the interactions while the fields evolve as if they were free. In this representation, the vacuum state evolves as
\be
|0\rangle_{\eta} = U(\eta,\eta_0) \ket
\ee
where $\ket$ is a free vacuum at some time $\eta_0$, assumed far in the past.
The two-point function becomes
\be
\brae \phi_{\vk_1}(\eta_1) U(\eta_1,\eta_2) \phi_{\vk_2}(\eta_2) \kete = \bra U(\eta_1,\eta_0)^{-1} \phi_{\vk_1}(\eta_1) U(\eta_1,\eta_2) \phi_{\vk_2}(\eta_2) U(\eta_2,\eta_0) \ket
\label{eqngenerale}
\ee
where the operator $U(\eta_1,\eta_0)^{-1}$ is just the inverse of the evolution operator $U(\eta_1,\eta_0)$, that is to say $U(\eta_0,\eta_1)$.

One can find an explicit representation the operator $U(\eta_1,\eta_2)$ which reads in the interaction picture
\be
U(\eta_1,\eta_2)=T \left[ {\rm e}^{ - i \int_{ \eta_2 }^{ \eta_1 } {\rm d}\eta' H_{I}(\eta') } \right]
\ee
where $H_I(\eta)$ is the interaction hamiltonian and $\eta_1$ is a conformal time greater than $\eta_2$.

One has to be careful with time ordering since, for $\eta_1<\eta_2$,
\be
U(\eta_2,\eta_1) = A \left [{\rm e}^{ + i \int_{ \eta_2 }^{ \eta_1 } {\rm d}\eta' H_{I}(\eta') } \right]
\ee
where A stands for the anti-time ordering operator.

\subsection{Effect of a small mass}

Let us study the correction to the non-massive scalar mode function from a potential of the form
\be
\label{quadratic}
U(\phi)=\frac{\delta m^2}{2}\phi^2
\ee
where $\delta m^2$ is assumed to be a small parameter.

As a first approximation, we can decompose the scalar field on a plane wave basis, taking into account the loop correction. The field then reads
\be
\chi(\vx,\eta) = \int \frac{ {\rm d}^3 \vk }{ (2\pi)^{3/2} } \left[ v_k(\eta) {\rm e}^{ i \vk \cdot \vx} a_{\vk} + v_k^*(\eta) {\rm e}^{- i \vk \cdot \vx} a_{\vk}^{\dagger}\right]
\ee
where we defined
\be
v_k(\eta)=u_k^{(0)}(\eta)+u_k^{(1)}(\eta)+\mathcal{O}(\delta m^4).
\ee
The mode function $u_k^{(0)}$ is the free mode function whereas $u_k^{(1)}$ is the first order correction in $\delta m^2$.

Expanding the mode function in power of the small parameter $\delta m^2$ 
in the equation of motion, one gets the system
\bea
u_k''^{(0)} + \left( k^2 - \frac{a''}{a} \right) u_k^{(0)} & = & 0 \\
u_k''^{(1)} + \left( k^2 - \frac{a''}{a} \right) u_k^{(1)} & = & - \delta m^2 a^2 u_k^{(0)}
\eea
that could be solved using the retarded Green function
\be
G^R(\eta,\eta')=- i \: \Theta(\eta-\eta') \left[ u_k^{(0)}(\eta) {u_k^{(0)}}^* (\eta') - {u_k^{(0)}}^* (\eta) u_k^{(0)}(\eta')\right]
\ee
where $\Theta$ is the Heaviside function. In the last expression, we used the normalisation (\ref{wronskien}).
To first order in $\delta m^2$, the correction to the mode function reads
\be
\label{Gfirstorder}
u_k^{(1)}(\eta) = -\int {\rm d}\eta' \delta m^2 a^2(\eta') u_k^{(0)}(\eta') G^R(\eta,\eta').
\ee
Expanding the Wightman function as
\be
\label{deltaG}
G_k(\eta_1,\eta_2) = u_k^{(0)}(\eta_1) { u_k^{(0)} }^*(\eta_2) + u_k^{(1)}(\eta_1) { u_k^{(0)} }^*(\eta_2) + u_k^{(0)}(\eta_1) { u_k^{(1)} }^*(\eta_2)+\mathcal{O}(\delta m^4)
\ee
and inserting (\ref{Gfirstorder}) into (\ref{deltaG}), one then obtains
\bea
\label{masseffect}
G_k(\eta_1,\eta_2) = G_k^{(0)}(\eta_1,\eta_2) & + & 2 \int_{-\infty}^{\eta_2} {\rm d}\eta'\delta m^2 a^2(\eta') {\rm Im} \left[ G_k^{(0)}(\eta_1,\eta') G_k^{(0)}(\eta_2,\eta') \right]\nonumber\\
& + & 2 \int_{\eta_2}^{\eta_1} {\rm d}\eta'\delta m^2 a^2(\eta') { G_k^{(0)} }^*(\eta_2,\eta') {\rm Im} \left[ G_k^{(0)}(\eta_1,\eta') \right].
\eea

This correction to the free propagator should be compared to the divergences which may be induced by other potentials
such as Yukawa couplings of the Wess-Zumino model.

\subsection{Radiative mass correction from a bosonic loop}

Let us now compute the one loop scalar self-mass induced by a quartic potential
\be
\label{quartic}
V(\phi)=\frac{\lambda}{4}\phi^4
\ee
with the related interaction Hamiltonian
\be
H_{I}(\eta)=\int {\rm d}^{3} \vx \sqrt{-g} \frac{\lambda}{4} \phi^4(x)=\int {\rm d}^{3} \vx \frac{\lambda}{4} \chi^4(x).
\ee
Expanding the evolution operator to first order in $\lambda$, we get from eq.~(\ref{eqngenerale})
\bea
\brae \chi_{\vk_1}(\eta_1) \chi_{\vk_2}(\eta_2) \kete
 = & - & i \int_{\eta_0}^{\eta_2} {\rm d} \eta' \bra \left[ \chi_{\vk_1}(\eta_1) \chi_{\vk_2}(\eta_2) , H_I(\eta') \right] \ket \nonumber\\
 & - & i \int_{\eta_2}^{\eta_1} {\rm d} \eta' \bra \left[ \chi_{\vk_1}(\eta_1) , H_I(\eta') \right] \chi_{\vk_2}(\eta_2) \ket.
\eea

Using the Wick theorem, this expression transforms into
\bea
&& \brae \chi_{\vk_1}(\eta_1) \chi_{\vk_2}(\eta_2) \kete \nonumber\\
& = & 6 \lambda \int_{\eta_0}^{\eta_2} {\rm d}\eta' \int {\rm d}^{3} \vx' {\rm Im} \left[ \bra \chi_{\vk_1}(\eta_1) \chi(x') \ket \bra \chi_{\vk_2}(\eta_2) \chi(x') \ket \bra \chi(x') \chi(x') \ket \right] \nonumber\\
& + & 6 \lambda \int_{\eta_2}^{\eta_1} {\rm d}\eta'\int {\rm d}^{3} \vx' {\rm Im} \left[ \bra \chi_{\vk_1}(\eta_1) \chi(x') \ket \right] \bra \chi(x') \chi_{\vk_2}(\eta_2) \ket \bra \chi(x') \chi(x') \ket.
\eea

Finally, using the definition (\ref{Gk}) one finds the contribution from the bosonic loop
\bea
\label{quarticeffect}
\brae \chi_{\vk_1}(\eta_1)\chi_{\vk_2}(\eta_2)\kete
& = & 6 \lambda \delta( \vk_1 + \vk_2 ) \int_{\eta_0}^{\eta_2} {\rm d}\eta' {\rm Im} \left[ G_k(\eta_1,\eta') G_k(\eta_2,\eta') \right] \int \frac{ {\rm d}^{3} \vp }{ (2\pi)^3 } G_p(\eta',\eta') \nonumber\\
& + & 6 \lambda \delta( \vk_1 + \vk_2 ) \int_{\eta_2}^{\eta_1} {\rm d}\eta'{\rm Im} \left[G_k(\eta_1,\eta') \right] G_k^*(\eta_2,\eta') \int \frac{ {\rm d}^{3} \vp }{ (2\pi)^3 } G_p(\eta',\eta').
\eea

Comparing eqs.(\ref{masseffect}) and (\ref{quarticeffect}), one finds a mass contribution from the quartic interaction at one loop order
\be
\label{bosonicmass}
\delta m^2_{\rm B} = \frac{3 \lambda}{ a^2(\eta') } \int \frac{ {\rm d}^{3} \vp }{ (2\pi)^3 } G_p(\eta',\eta')
\ee

This result holds in any spacially flat FLRW spacetime.

\subsection{Scalar self-mass from the fermionic loops}

We now pay attention to the contribution from fermionic loops through a Yukawa coupling of the form
\be
\label{Yukawa}
V(\phi,\psi)=\frac{\lambda}{\sqrt{2}}\overline{\psi}\phi\psi
\ee
where $\psi$ is a four component Majorana spinor.
The interaction Hamiltonian is given by
\be
H_{I}(\eta)=\int {\rm d}^{3} \vx \sqrt{-g} \frac{\lambda}{\sqrt{2}} \phi(x)\overline{\psi}(x)\psi(x).
\ee

The two-point function (\ref{eqngenerale}) involves three different evolution operators namely $U(\eta_2,\eta_0)$, $U^{-1}(\eta_1,\eta_0)$ and $U(\eta_1,\eta_2)$. The first non vanishing contribution - after the tree level - comes from the expansion of those operators to second order in the coupling constant. Expanding the first two as
\bea
U(\eta_2,\eta_0) &=& 1 - i \int_{\eta_0} ^ {\eta_2} {\rm d} \eta' H_{\rm I}(\eta') - \int_{\eta_0} ^ {\eta_2} {\rm d} \eta' \int_{\eta_0} ^ {\eta'} {\rm d} \eta'' H_{\rm I}(\eta') H_{\rm I}(\eta'') + \mathcal{O}(\lambda^3)\\
U^{-1}(\eta_1,\eta_0) &=& 1 + i \int_{\eta_0} ^ {\eta_1} {\rm d} \eta' H_{\rm I}(\eta') - \int_{\eta_0} ^ {\eta_1} {\rm d} \eta' \int_{\eta_0} ^ {\eta'} {\rm d} \eta'' H_{\rm I}(\eta'') H_{\rm I}(\eta') + \mathcal{O}(\lambda^3)
\eea
leads to three different terms in the two-point function (\ref{eqngenerale}) 
\bea
\label{U1}
&-& \int_{\eta_0}^{\eta_2} {\rm d}\eta' \int_{\eta_0}^{\eta'} {\rm d}\eta'' \bra \phi_{\vk_1}(\eta_1) \phi_{\vk_2}(\eta_2) H_{I}(\eta') H_{I}(\eta'') \ket \; , \\ 
&-& \int_{\eta_0}^{\eta_1} {\rm d}\eta' \int_{\eta_0}^{\eta'} {\rm d}\eta'' \bra H_{I}(\eta'') H_{I}(\eta') \phi_{\vk_1}(\eta_1) \phi_{\vk_2}(\eta_2) \ket \; , \\ 
&& \int_{\eta_0}^{\eta_1} {\rm d}\eta' \int_{\eta_0}^{\eta_2} {\rm d}\eta'' \bra H_{I}(\eta') \phi_{\vk_1}(\eta_1) \phi_{\vk_2}(\eta_2) H_{I}(\eta'') \ket
\eea
whereas including the expansion of $U(\eta_1,\eta_2)$ exhibits three other terms
\bea
& - & \int_{\eta_2}^{\eta_1} {\rm d}\eta' \int_{\eta_2}^{\eta'} {\rm d}\eta'' \bra \phi_{\vk_1}(\eta_1) H_{I}(\eta') H_{I}(\eta'') \phi_{\vk_2}(\eta_2) \ket \; , \\
&& \int_{\eta_0}^{\eta_1} {\rm d}\eta' \int_{\eta_2}^{\eta_1} {\rm d}\eta'' \bra H_{I}(\eta') \phi_{\vk_1}(\eta_1) H_{I}(\eta'') \phi_{\vk_2}(\eta_2) \ket \; , \\
& - & \int_{\eta_2}^{\eta_1} {\rm d}\eta'\int_{\eta_0}^{\eta_2} {\rm d}\eta'' \bra \phi_{\vk_1}(\eta_1) H_{I}(\eta') \phi_{\vk_2}(\eta_2) H_{I}(\eta'') \ket.
\label{U6}
\eea

All those terms describe fermionic loops and involve contractions over spinorial operators as well as integrations over running momenta. Such expressions - introduced in section II (see eqs.(\ref{Correlateur}) and (\ref{fermion3})) - only depend on the following quantity 
\be
f(\eta',\eta'';\Lambda,k) = \int \frac{ {\rm d}^3 \vp_1 }{ (2\pi)^3 } \int {\rm d}^3 \vp_2 \Delta_{p_1,p_2}(\eta',\eta'') \delta( \vp_1 + \vp_2 - \vk ),
\ee
where $\Delta_{p_1,p_2}(\eta',\eta'')$ is given by eq.(\ref{fermion3}) and $\Lambda$ is a ultraviolet three-dimensional cut-off. The integration in the infrared sector does not lead to any divergence as it could be expected from the spinors being conformal to Minkowski.
Summing all the terms (\ref{U1}-\ref{U6}) and noting that $f^*(\eta',\eta'';\Lambda,k)=f(\eta'',\eta';\Lambda,k)$, the two-point function may be written in a simpler form after cumbersome calculations
\bea
\label{PropLoopFermion}
\brae \chi_{\vk_1}(\eta_1) \chi_{\vk_2}(\eta_2) \kete
& = & 4 \lambda^{2} \delta^3( \vk_1 + \vk_2 ) \Bigg( \nonumber\\
&& \int_{\eta_0}^{\eta_2} {\rm d}\eta' \int_{\eta_0}^{\eta'} {\rm d}\eta''  \Big\{ {\rm Im} \left[ G_k(\eta_1,\eta') \right] {\rm Im} \left[ G_k(\eta_2,\eta'') f(\eta',\eta'';\Lambda,k_1) \right] \nonumber\\
& + & {\rm Im} \left[ G_k(\eta_2,\eta') \right] {\rm Im} \left[ G_k(\eta_1,\eta'') f(\eta',\eta'';\Lambda,k_1) \right] \Big\} \nonumber\\
& + & \int_{\eta_2}^{\eta_1} {\rm d}\eta'\int_{\eta_0}^{\eta'} {\rm d}\eta'' G_k^*(\eta_2,\eta'') {\rm Im} \left[G_k(\eta_1,\eta') \right] {\rm Im} \left[f(\eta',\eta'';\Lambda,k) \right)] \nonumber\\
& + & \int_{\eta_2}^{\eta_1} {\rm d}\eta' \int_{\eta_0}^{\eta2} {\rm d}\eta'' {\rm Im} \left[ G_k(\eta_1,\eta') \right] {\rm Im} \left[ G_k(\eta_2,\eta'') \right] f(\eta',\eta'';\Lambda,k) \Bigg).
\eea

As low running momenta do not contribute to $f(\eta',\eta'';\Lambda,k)$, there is no infrared contribution to
the bosonic masses.
Thus the radiative corrections come from the ultraviolet regime, i.e. from large momenta.
The expression (\ref{fermion3}) being a product of a divergent function by an oscillating one,
the main contribution occurs when the conformal times are close to each other. We can then
expand the expression (\ref{PropLoopFermion}) with respect to the infinitely small parameter
$\epsilon=\eta''-\eta'$ and the expansion of the Wightman function reads
\be
G_k(\eta,\eta'') = G_k(\eta,\eta') + (\eta''-\eta') G_k'(\eta,\eta') + \frac{ (\eta''-\eta')^2 }{2} G_k''(\eta,\eta') + \mathcal{O} \left[ (\eta''-\eta')^3 \right]
\ee
where $^{\prime}$ stands for a derivation with respect to the second variable.
We also define the real quantities indexed by $j$
\bea
I_j(\Lambda,k) & = & (-i) ^ {j-1} \int_{ \eta_0 }^{ \eta' } {\rm d}\eta'' \frac{ (\eta''-\eta')^j }{j!} \int \frac{ {\rm d}^3 \vp_1 }{ (2\pi)^3 } \int {\rm d}^3 \vp_2 \Delta_{p_1,p_2}(\eta',\eta'') \delta( \vp_1 + \vp_2 - \vk_1) \nonumber\\
& = & \int \frac{ {\rm d}^{3} \vp_1 }{ (2\pi)^3 } \int {\rm d}^{3} \vp_2 \frac{1}{ (p_1 + p_2)^{j+1}} \left[ 1 - \frac{ \vp_1 \cdot \vp_2 }{ p_1 p_2 } \right ] \delta( \vp_1 + \vp_2 - \vk )
\eea
to get the expansion
\be
\label{devlp}
\int_{\eta_0}^{\eta'} {\rm d}\eta'' G_k(\eta,\eta'') f(\eta',\eta'';\Lambda,k) = \sum_{j} i^{j-1} G_k^{(j)}(\eta,\eta') I_j(\Lambda,k).
\ee
The computations of $I_j(\Lambda,k)$ for $j=0,1,2$ are sufficient since $I_j(\Lambda,k)$ converges for $j>2$ and give
\bea
\label{I0}
I_0(\Lambda,k) & = & \frac{ \Lambda^2 }{ 4\pi^2 } - \frac{ k^2 }{ 8\pi^2 } \ln \left( \frac{ \Lambda }{ k } \right) + \mathcal{O} \left( \frac{k}{\Lambda} \right) \; , \\
\label{I1}
I_1(\Lambda,k) & = & \frac{1}{4 \pi^2} \Lambda +\mathcal{O} \left( \frac{k}{\Lambda} \right) \; ,  \\
\label{I2}
I_2(\Lambda,k) & = & \frac{1}{8 \pi^2} \ln \left( \frac{\Lambda}{k} \right) + \mathcal{O} \left( \frac{k}{\Lambda} \right).
\eea
Using the equation of motion for the Wightman function in the background metric
\be
G_k''(\eta,\eta') + k^2 G_k(\eta,\eta') = \frac{a''}{a} G_k(\eta,\eta')
\ee
and inserting eq.(\ref{devlp}) into eq.(\ref{PropLoopFermion}), the loop correction reads
\bea
\label{PropFermionInter}
\brae \chi_{\vk_1}(\eta_1) \chi_{\vk_2}(\eta_2) \kete
& = & -\frac{ \lambda^{2} }{ \pi^2 } \delta( \vk_1 + \vk_2 ) \Big \{ \nonumber\\
&& \int_{\eta_0}^{\eta_2} {\rm d}\eta' \left[ \Lambda^2 - \frac{a''}{2a} \ln{ \left( \frac{ \Lambda }{ k } \right) } \right] {\rm Im} \left[ G_{k_1}(\eta_1,\eta') G_{k_1}(\eta_2,\eta') \right] \nonumber\\
& + & \int_{\eta_2}^{\eta_1} {\rm d}\eta' \left[ \Lambda^2 - \frac{a''}{2a} \ln{ \left( \frac{ \Lambda }{ k } \right) } \right] G_{k_1}^*(\eta_2,\eta') {\rm Im} \left[ G_{k_1}(\eta_1,\eta') \right] \Big\} \nonumber\\
& - & \frac{ \lambda^2 }{ 4 \pi^2 } {\rm Re} \left[ G_k(\eta_1,\eta_2) \right] \ln{ \left( \frac{ \Lambda }{ k } \right) } \nonumber\\
& + & 4 \lambda^2 \int_{\eta_2}^{\eta_1} {\rm d}\eta'\int_{\eta_0}^{\eta2} {\rm d}\eta'' {\rm Im} \left[ G_k(\eta_1,\eta') \right] {\rm Im} \left[ G_k(\eta_2,\eta'') \right] f(\eta',\eta'';\Lambda,k).
\eea
Let us compute the last term. Its main contribution comes from $\eta''$ and $\eta'$ being close to $\eta_2$. We can then expand the Green function with $\eta'-\eta_2\rightarrow0$ and $\eta''-\eta_2\rightarrow0$
\be
{\rm Im} \left[ G_k(\eta_1,\eta') \right] {\rm Im} \left[ G_k(\eta_2,\eta'') \right] = \sum_{j,l} {\rm Im} \left[ G_k^{(j)}(\eta_1,\eta_2) \right] {\rm Im} \left[ G_k^{(l)}(\eta_2,\eta_2) \right] \frac{ (\eta'-\eta_2)^j }{ j! } \frac{ (\eta''-\eta_2)^l }{ l! }.
\ee
We are left with two terms of the form
\bea
\int_{\eta_0}^{\eta2} {\rm d}\eta'' \frac{ (\eta''-\eta_2)^j }{ j! } {\rm e}^{- i ( p_1 + p_2 )( \eta_2 - \eta'' ) } & = & \frac{ i^{j-1} }{ ( p_1 + p_2 )^{j+1} } \; ,\\
\label{approxT12}
\int_{\eta_2}^{\eta1} {\rm d}\eta'' \frac{ ( \eta' - \eta_2 )^j }{ j! } {\rm e}^{- i ( p_1 + p_2 ) ( \eta' - \eta_2 ) } & = & \frac{ (-i)^{j-1} }{ ( p_1 + p_2 )^{j+1} } \left[ {\rm e}^{- i ( p_1 + p_2 )( \eta_2 - \eta_1 ) } - 1 \right].
\eea
The equation (\ref{approxT12}) provides us with a term whose phase is proportional to the difference
of the conformal times. Since this term oscillates rapidly when $|\eta_1-\eta_2|>1/\Lambda$,
we can neglect its ultraviolet contribution unless the times are equal, which would lead to a
trivial result for (\ref{approxT12}).

The last term of eq. (\ref{PropFermionInter}) then becomes
\bea
&& \int_{\eta_2}^{\eta_1} {\rm d}\eta' \int_{\eta_0}^{\eta2} {\rm d}\eta'' {\rm Im} \left[ G_k(\eta_1,\eta') \right] {\rm Im} \left[ G_k(\eta_2,\eta'') \right] f(\eta',\eta'';\Lambda,k) \nonumber\\
& = & \sum_{j,l} {\rm Im} \left[ G_k^{(j)}(\eta_1,\eta_2) \right] {\rm Im} \left[ G_k^{(l)} (\eta_2,\eta_2) \right] (i)^{l-j} I_{l+j+1} (\Lambda,k).
\eea
The only divergent term is obtained when $l=1$ and $j=0$ and reads
\be
- i\, {\rm Im} \left[ G_k(\eta_1,\eta_2) \right] {\rm Im} \left[ G_k'(\eta_2,\eta_2) \right] I_2(\Lambda,k) = - \frac{i}{2} {\rm Im} \left[ G_k(\eta_1,\eta_2) \right] I_2(\Lambda,k)
\ee
where we used eq. (\ref{wronskien}).

Finally, we are left with
\bea
\label{PropFermion}
\brae \chi_{\vk_1}(\eta_1) \chi_{\vk_2}(\eta_2) \kete
&=& -\frac{ \lambda^{2} }{ \pi^2 } \delta( \vk_1 + \vk_2 ) \Bigg \{ \nonumber\\
&& \int_{\eta_0}^{\eta_2} {\rm d}\eta' \left[ \Lambda^2 - \frac{a''}{2a} \ln{ \left( \frac{ \Lambda }{ k } \right) } \right] {\rm Im} \left[ G_{k_1}(\eta_1,\eta') G_{k_1}(\eta_2,\eta') \right] \nonumber\\
& + & \int_{\eta_2}^{\eta_1} {\rm d}\eta' \left[ \Lambda^2 - \frac{a''}{2a} \ln{ \left( \frac{ \Lambda }{ k } \right) } \right] G_{k_1}^*(\eta_2,\eta') {\rm Im} \left[ G_{k_1}(\eta_1,\eta') \right] \nonumber\\
& - & \frac{1}{4} G_k(\eta_1,\eta_2) \ln{ \left( \frac{ \Lambda}{k} \right) } \Bigg\}.
\eea

There is no linear divergence left, in agreement with the local Lorentz invariance. The additional term may be cancelled by a field strength renormalisation setting
\be
v_k(\eta)\rightarrow(1+\delta Z)v_k(\eta).
\ee
with
\be
\label{dZ}
\delta Z = - \frac{ \lambda^2 }{ 8\pi^2 } \ln{ \left( \frac{ \Lambda }{ \mu } \right) }
\ee
where $\mu$ is a typical energy scale depending on the renormalisation condition.
>From eq.(\ref{masseffect}), one can see that the fermionic loop gives a mass to the scalar field
\be
\label{fermionicmass}
\delta m^2_{\rm F} = - \frac{ \lambda^2 }{ 2 \pi^2 a^2(\eta)} \left[ \Lambda^2 - \frac{a''}{2a} \ln \left( \frac{ \Lambda }{ \mu } \right) \right].
\ee

One can remark a background dependent logarithmic divergence proportional to the scalar curvature.

In the case of a Yukawa coupling (\ref{Yukawa}), the propagator of fermions is just modified by a trace over an odd number of gamma matrices which vanishes. Fermions do not acquire a mass through a Yukawa coupling.

\section{Interpretation}

In the context of the Lagrangian (\ref{LWZ}), the one loop scalar masses read from eqs.(\ref{bosonicmass}) and (\ref{fermionicmass})
\be
\label{netmass}
\delta m^2 = \frac{ \lambda^2 }{ 2 \pi^2 a^2(\eta) } \left[ 4 \int {\rm d}p p^2 G_p(\eta,\eta) - \Lambda^2 + \frac{a''}{2a} \ln \left( \frac{ \Lambda }{ \mu } \right) \right].
\ee
In the following, we compute this expression for different FLRW spacetimes and special attention is paid to de Sitter spacetime.

\subsection{Flat spacetime}

In the flat case, the exact computation of the radiative mass corrections in the massive case may be easily dealt with and one expects to recover the Minkowski results by setting $a=1$ and
\be
G_k(\eta,\eta) = \frac{1}{ 2 \omega_k } = \frac{1}{ 2 \sqrt{ k^2 + m^2 } }.
\ee
Considering a non-vanishing mass entails an additional correction from the trilinear coupling in the bosonic sector.
Finally, one finds a zero mass correction for both the scalar fields and a field strength renormalisation as it
should be expected from a supersymmetric model in Minkowski spacetime \cite{Renorm,SUGRA1}.

\subsection{Generic flat FLRW spacetime}

To compute the expression (\ref{netmass}) in a generic flat FLRW spacetime, we need to specify a vacuum since the choice of a vacuum state is not imposed \cite{birrel}. The simplest choice consists in a Bunch-Davies vacuum whose limit is Minkowskian far in the past. With this convention, for any spatially flat FLRW spacetime with constant acceleration, the mode function is given by \cite{Wood}
\be
\label{FLRWMode}
u_k(\eta) = \sqrt{ \frac{ \pi r x^r }{ 4 k } } x^\alpha H_\nu^{(1)}(rx)
\ee
where
\be
\label{def}
y = - \frac{1}{ a H^2 } \frac{ {\rm d}H }{ {\rm d}\eta } = - \frac{1}{H^2} \frac{ {\rm d}H }{ {\rm d}t } \: , \: \: r = \frac{1}{ 1-y } \: , \: \: \nu = r + \frac{1}{2} \: , \: \: \alpha = - \frac{r}{2} y \: , \: \: x = \frac{k}{ a H }
\ee
and where $H_\nu^{(1)}$ is the Hankel function of the first kind with the index $\nu$. We will only pay attention to period with constant acceleration, i.e. $y = {\rm cst}$.
To study the ultraviolet divergences, we shall use the following expansion (see e.g. \cite{gradshteyn})
\be
\label{UVExpand}
H_{\nu}^{(1)}(z)=\sqrt{\frac{2}{\pi z}}{\rm e}^{i(z-\nu\frac{\pi}{2}-\frac{\pi}{4})}\sum_{n=0}^\infty\frac{\Gamma(\nu+n+\frac{1}{2})}{n!\Gamma(\nu-n+\frac{1}{2})}\left(\frac{i}{2z}\right)^n.
\ee
Expanding the expression (\ref{FLRWMode}) for large momenta, one gets
\be
u_k(\eta)=\frac{1}{\sqrt{2k}}{\rm e}^{ i ( r x - r \pi / 2 - \pi / 2 ) } \left[ 1 + i \frac{ r + 1 }{ 2x } - \frac{ ( r+2 ) ( r^2-1 )}{ 8 r x^2 } \right] + \mathcal{O} \left( \frac{1}{x^3} \right)
\ee
which leads to
\be
G_k ( \eta , \eta) = u_k(\eta) u_k^*(\eta) = \frac{1}{2k} \left( 1 + \frac{ r+1 }{ 2rx^2} \right) + \mathcal{O} \left( \frac{1}{x^4} \right) = \frac{1}{2k} \left( 1 + \frac{a''}{2a} \frac{1}{k^2} \right) + \mathcal{O} \left( \frac{1}{x^4} \right).
\ee
Inserting this result into (\ref{netmass}), one obtains the one loop mass correction in any constant acceleration FLRW spacetime in the ultraviolet regime
\be
\delta m^2 = \delta m^2_{\rm B} + \delta m^2_{\rm F} = \frac{ 3 \lambda^2 }{ 4 \pi^2 } \frac{a''}{a^3} \ln \left( \frac{ \Lambda^2 }{ \mu^2 } \right) + \mathcal{O} \left( \frac{ \mu }{ \Lambda } \right)
\ee
where $\mu$ is a typical energy scale.

There is no quadratic divergence left since we should recover a supersymmetric action on small scales. The local Lorentz invariance prevents any linear divergence. The only remaining one is due to the curvature which is in agreement with (\ref{meff}) and follows directly from a WKB approximation from the equation of motion (\ref{KG}).

The infrared regime may also lead to divergences. We will only pay attention to the infrared divergence
due to the bosonic loops since low momenta from the fermionic loops have no contribution.
Using the following expansion in the case of a non-integer index~\cite{gradshteyn}
\be
\label{IRExpand}
H_{\nu}^{(1)}(z)\sim\frac{1}{\Gamma(1+\nu)}\left(\frac{z}{2}\right)^{\nu}+\frac{i}{\sin(\pi\nu)}\frac{1}{\Gamma(1-\nu)}\left(\frac{z}{2}\right)^{-\nu} \; ,
\ee
the Wightmann function reads in the small momentum limit
\be
G_k(\eta,\eta) \sim \frac{ \pi r }{ 4 a H } \left[ \frac{ (r/2) ^ {2 \nu } }{ \Gamma^2 (1 + \nu) } ( \frac{k}{ a H } )^{ 2 \alpha + r - 1 + 2 \nu } + \frac{1}{ \sin^2( \pi \nu) } \frac{ (r/2)^{- 2 \nu } }{ \Gamma (1 - \nu) } ( \frac{k}{ a H } )^{ 2 \alpha + r - 1 - 2 \nu} \right].
\ee
Inserting this expansion into eq. (\ref{bosonicmass}) gives a vanishing contribution if both the integrals converge, i.e. if
\bea
2\alpha+r+1+2\nu > -1\nonumber\  \  \ {\rm and}\  \ \
2\alpha+r+1-2\nu > -1.\eea 
Hence, in the massless
case, there are infrared divergences for any $|\nu| > 3/2$. For a
scale factor growing as $ a(t) \sim t^ {\beta}$, we get from
eq.(\ref{def}) \be \nu = \frac{ 3 \beta - 1 }{ 2 \beta - 2 }. \ee
For power law expansion such as $ \beta > 2/3 $, the scalar mass
gets an infrared divergent contribution. In the limit of a
massless scalar field in a de Sitter space, $\nu$ exactly equals
$3/2$ and an infrared logarithmic divergence does appear. This
infrared behavior has been investigated in~\cite{infrarouge}.

\subsection{Consequences on de Sitter inflation}

In this subsection, we apply the previous results to the specific case of de Sitter spacetime which is assumed to properly describe the phase of inflation in the early universe. In particular, light scalar fields are bound to generate significant fluctuations during inflation if their masses are small compared to the Hubble scale. We want to check that their masses are protected against one loop mass corrections.

In the following, we shall consider a coupling to the curvature which may be added to the Lagrangian (\ref{LWZ}).
The curvature in a de Sitter spacetime is constant and given by $R=12H^2$. Hence, the additional potential
looks like a mass term for the scalars
\be
V(\phi) = \frac{1}{2} \xi R ( \phi_1^2 + \phi_2^2 ).
\ee

Using the previous results, the effective masses found in (\ref{meff}) read
\be
m_{\rm eff}^2 \simeq - \frac{a''}{a^3} + \frac{1}{2} \xi R = 2 ( 6 \xi - 1 ) H^2.
\ee
This effective mass is constant and vanishes in the case of conformally coupled scalar fields.
To express the quantum corrections to the masses, let us choose the Bunch-Davies vacuum \cite{birrel} for which the mode function reads
\be
u_k(\eta) = \sqrt{ \frac{ \pi }{ 4k } } \sqrt{ - k \eta } H_{\nu}^{(1)}(- k \eta )
\ee
where $H_{\nu}^{(1)}$ is the Hankel function of the first kind and $\nu^2=9/4-12\xi$.
Applying the previous expansions of Hankel function, the self mass in the ultraviolet regime reads
\be
\delta m^2 = \frac{ 3 \lambda^2 H^2 }{ 2 \pi^2 } ( 1 - 6 \xi ) \ln \left( \frac{ \Lambda }{ \mu } \right).
\ee

Not surprisingly, there is no additional ultraviolet logarithmic divergence
for a conformally coupled scalar field with~$\xi=1/6$.

The de Sitter solutions for the massive mode functions are well known and read
in the Bunch-Davies vacuum
\be
u_k(\eta)=\sqrt{ \frac{ \pi }{ 4k } } \sqrt{ - k \eta } H_{\nu'}^{(1)}(- k \eta )
\ee
with $\nu '^2=9/4-m^2/H^2-12\xi$. From eqs.~(\ref{bosonicmass}) 
and~(\ref{IRExpand}), we can see
that the contribution to the mass from the infrared regime
vanishes for any index such as $ | {\rm Re} (\nu') | < 3/2 $, i.e. as soon as
$m^2+12\xi H^2>0$.

These results may be
applied to the minimally coupled case for which the Wigthman
function takes the simpler form \be G_k(\eta,\eta) = \frac{1}{2k}
( 1 + \frac{1}{ k^2 \eta^2 } ). \ee The radiative mass corrections
read \bea \label{cutoff}
\delta m^2 _{\rm B} &=& \frac{ \lambda^2 }{ 2 \pi^2 a^2 } \left[ \Lambda^2 + 2 a^2 H^2 \ln \left( \frac{\Lambda}{\Lambda_{\rm IR} } \right) \right] \; , \\
\delta m^2 _{\rm F} &=& - \frac{ \lambda^2 }{ 2 \pi^2 a^2 } \left[ \Lambda^2 -  a^2 H^2  \ln \left( \frac{ \Lambda }{ \mu } \right) \right] \; , \\
\delta m^2 &=& \delta m^2 _{\rm B} + \delta m^2 _{\rm F} = \frac{ 3 \lambda^2 H^2 }{ 2 \pi^2 } \ln \left( \frac{ \Lambda }{ \mu } \right) + \frac{ \lambda^2 H^2 }{ \pi^2 } \ln \left( \frac{ \mu }{ \Lambda_{\rm IR} } \right).
\label{mdeSitter}
\eea

For the light scalar fields to develop a significant fluctuations
one should get rid of quadratic divergences - as it is the case
in supersymmetric models - and impose a condition to the coupling
constant such as the mass is protected against radiative
correction \be \lambda^2 \ln \left( \frac{ \Lambda }{ \mu }
\right) \ll 1 \ee which could be easily satisfied for a coupling
constant smaller than $10^{-2}$.

The infrared divergence that appears in the mass corrections is
related to the large scale structure of spacetime and have been
extensively discussed in the literature
\cite{VilFord, Linde, birrel, Staro3, Wood, woodard1, woodard2, woodard3, woodard4}. Such a divergence naturally
appears in the real space propagator, the cut-off being of order of
the initial comoving horizon, i.e. : \be \Lambda_{\rm IR} \simeq
a_0 H, \ee where $a_0$ is the scale factor at some initial time
\footnote{The correction to the mass from the bosonic loop has
also been computed in another way in \cite{woodard4}. These two
results are compatible in the only case of a time dependent
cut-off growing as the scale factor, i.e. being proportional to
the comoving horizon in a de Sitter space.}. In the context of de
Sitter spacetime, the transition between a quantum and a classical
behavior occurs at the horizon crossing : perturbations modes
leave the horizon and behave classically outside the horizon
\cite{BU1,Staro1,Staro2}. It is then natural to pay attention to a
classical description of the divergences which is the aim of the
following subsection.

\subsection{Connection to a classical viewpoint}

During inflation, the universe exponentially grows. The equation of evolution for a stochastic classical field which describes the large scale perturbations is given by eq.(\ref{KG1}) in the low momentum limit
\be
\label{KGclassique}
\delta s'' + 2 H \delta s' = - a^2 S( \delta s )
\ee
where $S$ stands for the source term. In the case of a parabolic potential (\ref{quadratic}) with a quartic self interaction (\ref{quartic}), it takes the form
\be
S(\delta s)=\delta m^2 \delta s+\lambda \delta s^3.
\ee
In a pertubative approach, we can expand the field into powers of the coupling constant
\be
\delta s = \delta s^{(0)} + \delta s^{(1)} + ...
\ee
and define the Fourier transform of the field for each order of perturbation
\be
u_{\vk}^{(n)}(\eta)=\int\frac{{\rm d}^3 \vk }{(2\pi)^{3/2}}\delta s^{(n)}(\vx,\eta){\rm e}^{-i\vk\cdot\vx}.
\ee
As a stochastic gaussian field, $u_{\vk}^{(0)}$ has the following properties
\bea
\label{property1}
{u_{\vk}^{(0)}}^* & = & u_{-\vk} ^{(0)} \\
\label{property2}
\langle u_{\vk}^{(0)} u_{\vk'}^{(0)} \rangle & = & \delta^3 ( \vk + \vk' ) P(k)
\eea
where the brackets stand for an ensemble average and the power spectrum, $P(k)$, entirely defines the field distribution.
To zeroth order, the modes are frozen and we get from the equation of motion
\be
\delta s^{(0)} = {\rm cst}.
\ee
The first order is sourced by the free field
\be
{\delta s^{(1)}}'' + 2 H { \delta s^{(1)} }' = - a^2 S( \delta s^{(0)} )
\ee
where all the time dependence in the r.h.s. is expressed in the factor $a^2$.
We then get
\be
{ \delta s^{(1)} }'(\vx,\eta_1) = - S( \delta s^{(0)} ) {\rm e}^{- \frac{2}{ a(\eta_1) } } \int_{\eta_0}^{\eta_1} {\rm d}\eta_2 \: a(\eta_2)^2 {\rm e}^{ \frac{2}{ a(\eta_2) } }.
\ee

In the case of a parabolic potential (\ref{quadratic}), the first order mode function becomes in momentum space
\be
\label{ClassicOrderK1}
u_{\vk}^{(1)}(\eta) = -\delta m^2 I(\eta) u_{\vk}^{(0)}
\ee
where we defined
\be
I(\eta)=\int_{\eta_0}^{\eta} {\rm d}\eta_1 {\rm e}^{- \frac{2}{ a(\eta_1) } } \int_{\eta_0}^{\eta_1} {\rm d}\eta' \: a(\eta')^2 {\rm e}^{ \frac{2}{ a(\eta') } }.
\ee

Inserting eqs.(\ref{property1}), (\ref{property2}) and (\ref{ClassicOrderK1}) into (\ref{deltaG}), one gets
\be
\langle u_{\vk}(\eta_1) u_{\vk'}(\eta_2) \rangle = \delta^3( \vk + \vk' ) P(k) - \delta m^2 \delta^3( \vk + \vk' ) P(k) \left[ I(\eta_1) + I(\eta_2) \right] \int \frac{ {\rm d}^3 \vp }{ (2 \pi)^3 } P(p).
\ee

In the case of a quartic potential (\ref{quartic}), one finds in momentum space
\be
\label{ClassicOrderK2}
u_{\vk}^{(1)}(\eta) = - \lambda I(\eta) \int \frac{ {\rm d}^3 \vk_1 }{ ( 2 \pi )^{ 3/2 } } \int \frac{ {\rm d}^3 \vk_2 }{ ( 2 \pi )^{ 3/2 } } \int {\rm d}^3k_3 \delta^3 ( \vk_1 + \vk_2 + \vk_3 - \vk) u_{\vk_1}^{(0)} u_{\vk_2}^{(0)} u_{\vk_3}^{(0)}
\ee
and the two-point function becomes
\be
\langle u_{\vk}(\eta_1) u_{\vk'}(\eta_2) \rangle = \delta^3 ( \vk + \vk' ) P(k) - 3 \lambda \delta^3( \vk + \vk' ) P(k) \left[ I(\eta_1) + I(\eta_2) \right] \int \frac{ {\rm d}^3 \vp }{ ( 2 \pi )^3 } P(p).
\ee

The correction to the mass from the quartic interaction reads
\be
\delta m^2 = 3 \lambda \int \frac{ {\rm d}^3 \vk }{ ( 2 \pi )^3 } P(k)
\ee
which is the same result as in the quantum level calculation since
\be
P(k) =\frac{1}{a^2} G_k(\eta,\eta)
\ee
with $k$ going to zero. In the case of a Harrison-Zeldovitch spectrum, i.e. $P(k)\sim k^{-3}$, one recovers the same infrared logarithmic divergence as in the previous subsection.

\section{Conclusion}

We calculated the radiative corrections to the masses of test
scalar and fermionic fields in a Wess-Zumino toy model embedded in
a spatially flat FLRW spacetime. Our results emphasized that the
curvature acts as an effective time dependent mass $m^2_{\rm eff} =
- R/6 $ for bosons whereas fermions stay massless. In a de Sitter
spacetime, this effective mass is constant and given by the Hubble
scale.

The final results show that radiative corrections exhibit
divergent contributions to the scalar masses both in the
ultraviolet and infrared regimes. The leading ultraviolet
divergences, quadratic and linear in the energy cutoff, cancelled
both as expected since one should recover the Minkowskian
structure in any locally inertial frame. However a subleading
ultraviolet divergence, logarithmic in the energy cutoff and
proportional to the spacetime curvature, survives. If it gives
masses to scalar fields, it does not however render them too heavy
compared to the Hubble scale. Hence, no significant fine tuning
will be required for those fields to develop superhorizon
fluctuations. Moreover, the infrared divergences that we observe
arise as soon as the scale factor grows faster than $t^{2/3}$.
They can be interpreted as those due to a self-interacting
stochastic classical field.

These results call for some comments.

First, one should carefully deal with supersymmetric models in an
expanding universe. In supergravity theories, the behavior of the
fields - including the graviton and the gravitino - has to be
deduced from an action which is locally supersymmetric. This task
is often difficult to handle with. Imposing an expansion of the
background, i.e. a dynamics for the graviton, and setting the
vacuum expectation values of the fermions to zero would generally
lead to a supersymmetry breaking as confirmed by our result.

In this context, the use of the Coleman-Weinberg formula in case
of an expanding universe can be questioned. For instance, in
inflationary models - such as F or D-term inflation models - the
one loop effective terms due to the curvature should be compared
to the usual one loop terms of the flat spacetime limit~\cite{CW}.
Although one does not expect this contribution to deeply affect
the dynamics at one loop order, consequences of such terms should
probably be studied in more details~\cite{CWcourbe, CWcourbe2}.

Finally what we have shown here is that light test scalar fields
do not acquire a contribution larger than the Hubble constant from
one loop corrections when they are embedded in chiral multiplet.
This result extents what was known for the inflaton field. As a
consequence it is reasonable to assume that light scalar fields
more or less generically develop superhorizon isocurvature
fluctuations. And then, as was shown in ~\cite{BU1}, nothing
prevents those from containing significant non-Gaussianities although
whether it is generically the case is still an open question. It
anyway puts non-Gaussian inflationary models on a much firmer
ground.

\vskip0.5cm
{\bf Acknowlegment:}
We want to thank J. Mourad, R.P. Woodard, J. Iliopoulos, N.
Chatillon, P. Brax, J. Rocher for fruitful discussions.

%%%%%%%%%%%%%%%%%%%%%%%%%%%%%%%%%%%%%%%%%%%%%%%%%%%%%%%%%%%%%%%%%%%

\end{document}